\documentclass[aps,pra,superscriptaddress,showpacs,floatfix]{revtex4}
\usepackage{graphicx}
\begin{document}


\title{Entanglement sharing in one-particle states}

\author{Arul Lakshminarayan}
\email[]{arul@prl.ernet.in}
\homepage[]{http://www.prl.res.in/~arul}
\affiliation{Department of Physics, Indian Institute of Technology, 
Kanpur, 208016, India.}
\affiliation{Physical Research Laboratory,
Navrangpura, Ahmedabad, 380009, India.}
\author{V. Subrahmanyam}
\email[]{vmani@iitk.ac.in}
\affiliation{Department of Physics, Indian Institute of Technology, 
Kanpur, 208016, India.}

\date{\today}

\begin{abstract}
Entanglement sharing among sites of one-particle states is
considered using the measure of concurrence. These are the
simplest in an hierarchy of number-specific states of many qubits
and corresponds to ``one-magnon'' states of spins. We study the
effects of onsite potentials that are both integrable and nonintegrable.
In the integrable case we point to a metal-insulator transition that 
reflects on the way entanglement is shared.
In the nonintegrable case
the average entanglement content 
increases and saturates along with a transition to classical chaos.
Such quantum chaotic states are 
shown to have universal concurrence distributions that
are modified Bessel functions derivable within random matrix theory.
Time-reversal breaking and time evolving states are shown to 
possess significantly higher entanglement sharing capacity that 
eigenstates of time-reversal symmetric systems. We use the ordinary 
Harper and kicked Harper Hamiltonians as model systems.
\end{abstract}

\pacs{03.67.Mn,05.45.Mt}

\maketitle
\newcommand{\newc}{\newcommand}
\newc{\beq}{\begin{equation}}
\newc{\eeq}{\end{equation}}
\newc{\kt}{\rangle}
\newc{\br}{\langle}
\newc{\beqa}{\begin{eqnarray}}
\newc{\eeqa}{\end{eqnarray}}
\newc{\pr}{\prime}
\newc{\longra}{\longrightarrow}
\newc{\ot}{\otimes}
\newc{\p}{\partial}
\newc{\rarrow}{\rightarrow}
\newc{\h}{\hat}
\newc{\f}{\frac}
\newc{\al}{\alpha}
\newcommand{\dg}{\dagger}

\section{Introduction}

Entanglement is a property of quantum systems that sets it apart from
those that are classical. Although it has been recognized as such from
the early days of quantum mechanics, a spurt of understanding
entangled states, both mathematically and generating them
experimentally, has occurred in the past few years. Due to its
potential as a resource in various tasks of quantum information
processing it has moved from philosophical debates to the center stage
of a large body of concrete work. For a recent review of the ideas
involved we refer to \cite{DagmarBrusJMP02}.

Entanglement within pure states of a bipartite system can be measured
by the von Neumann entropy of the reduced density matrices. For a
mixed state, while the entanglement can be measured as the average
entanglement of its pure-state decomposition, the existence of an
infinite number of such decompositions makes their minimization over
this set a nontrivial task. Wootters and Hill \cite{WootPRL9899} carried out
such a procedure for the case of two two-state (qubit) systems and
showed that a new quantity they called concurrence was a measure of
entanglement. This facilitated the study of entanglement sharing among
many qubits. One view of quantum entanglement, as a correlation that
is much stronger than any that is classical, is borne out here as two
maximally entangled qubits cannot be entangled with any other, they
will necessarily have to give up some of their correlation in order to
share it with a third. At this stage the nature of entanglement
sharing among many qubits is being studied intensively. Results are
known for specific subsets of states in the many qubit Hilbert
spaces \cite{Woot}. Recent work has explored entanglement sharing 
among higher state (higher than qubits) systems \cite{Woot02}.

Due to the possibility of using spins as qubits in quantum computers,
there have been many studies of the eigenstates of well known spin
Hamiltonians such as the Heisenberg model, Ising model in a transverse
field etc. \cite{Woot,Sougato}, and related itinerant fermion systems
\cite{Zanardi}. There has been a conjecture that for complex quantum
systems, entanglement will be an indicator of quantum phase
transitions \cite{Nielsen,Nature}. While these latter works have
explored complexity from the viewpoint of many particle,
thermodynamic, systems, few particle systems that are classically
chaotic are also complex in their own way with well-studied spectral
transitions occurring in the quantum systems \cite{Gutz,Haake}.  For
bipartite systems of this kind pure-state entanglement has been shown
to be sensitive to the presence of classical chaos and the typical
value of entanglement has been calculated from random matrix theory
(RMT), including the distribution of the eigenvalues of the reduced
density matrices \cite{Arul01,JayArul02,Zyck}.

In this paper we study states in the simplest subspace of the
$2^N$-dimensional Hilbert space of $N$ qubits, the $N$-dimensional
subspace spanned by vectors with only one qubit in a different state
from the rest, in some fixed single qubit basis. These are the
``single-particle states'' within which we calculate entanglement
sharing amongst the $N$ qubits. Thus we think of a one-dimensional
chain of $N$ sites with a single particle hopping among these. The
entanglement among the qubits is then the entanglement among the sites
themselves.  We will use the (spinless) fermion language as the
connection between the fermion operators and the spin-half algebra of
Pauli matrices is established through the Jordan-Wigner transformation
\cite{JW}. Although, we will not need to use these here due to our
restriction to single-particle states, the extension to higher number
of particles is then straightforward.

In the integrable case, we show how the onsite potential can 
decrease the average entanglement present in a state and point 
to a sharp fall that can be identified in the Harper Hamiltonian to
a metal-insulator transition.
In the nonintegrable case we show that the average entanglement content 
increases and saturates
along with a classical transition to complete chaos. Simultaneously
near neighbor entanglement gets destroyed and distant qubits start to
get significantly entangled. The effect of time-reversal symmetry breaking
is significant and leads to a larger entanglement content in the state.
Random matrix theory is then used to explain these features and is 
shown to be successful in predicted the distribution of concurrence
in an ensemble of chaotic states. 

\section{Preliminaries}
In this section we collect results that set the formalism and
notation. For completeness we first recall the definition of
concurrence. Consider a bipartite composite system of subsystems $A$
and $B$. A pure-state $|\psi\kt$ of the composite system is separable
if it can be written as an outer product of states from $A$ and
$B$. In general this is possible only if the reduced density matrices,
after tracing out either $A$ or $B$ is itself a pure-state density
matrix. Thus the von Neumann entropy of the reduced density matrices
is a measure of how entangled the pure-state $|\psi\kt$ is, and this
is also the Shannon entropy of the state in the Schmidt decomposed
form \cite{PeresPreskill}. The entanglement $E(|\psi\kt)$ in this
case is defined to be the von Neumann entropy of the subsytems
described by $\rho_A$ or $\rho_B$, which are obtained by tracing out
the states corresponding to subsystems $B$ and $A$ respectively:
\beq
\label{basicdef1}
E(|\psi\kt)=-\mbox{tr}(\rho_A \log_2(\rho_A))=-\mbox{tr}(\rho_B\log_2(\rho_B)).
\eeq

For bipartite density matrices the measures of entanglement are not so
easily calculable. The entanglement of formation for a general state
$\rho^{AB}$ is defined in the following way. For a given 
decomposition of the mixed state in terms of ensembles of pure-states 
$|\psi_i\kt$ specified with probabilities $p_i$:
\beq
\label{basicdef2}
\rho^{AB}=\sum_{i} p_i |\psi_i\kt \br \psi_i|,
\eeq
one may find the average entanglement present in all the pure states
involved.  The entanglement of formation is then defined as the
minimum of this average over all such possible pure-state decompositions:
\beq
\label{basicdef3}
E(\rho^{AB})=\mbox{min}\;\; \sum_{i} p_i E(|\psi_i\kt).
\eeq
This is one of the measures of entanglement, and is called the
entanglement of formation as it refers to the optimal ability to form
such mixed states from maximally entangled pure states using only
local operations on subsystems $A$ and $B$ and classical communication
between them \cite{BBPS}. For a general bipartite mixed state no
explicit equation is known for this quantity.  For a pair of qubits
Wootters \cite{WootPRL9899} found such an expression that enables one
to calculate the entanglement of formation from a knowledge of
$\rho^{AB}$, which we recall for completeness.

Defining a spin-flip operator, which takes $\rho^{AB}\equiv \rho$ to 
\beq
\tilde{\rho}=(\sigma_y\otimes\sigma_y)\rho^{*} (\sigma_y\otimes\sigma_y),
\eeq
the concurrence of $\rho$ is defined to be:
\beq
C(\rho)=\mbox{max}\;\;\{\sqrt{\lambda_1}-\sqrt{\lambda_2}-\sqrt{\lambda_3}-\sqrt{\lambda_4},\,0\}
\eeq
where $\lambda_i$ are the eigenvalues of the non-Hermitian matrix
$\rho \tilde{\rho}$. Wootters \cite{WootPRL9899} showed that the
entanglement of formation of $\rho^{AB}$ is a monotonic function of
its concurrence and that as the concurrence varies over its possible
range $[0,1]$, the entanglement of formation also varies from 0 to 1,
thus concurrence is itself a good measure of entanglement.

For eigenstates of the number operator, as we will consider in this
paper, the reduced density matrix of two sites has a special form that
has already been studied and exploited in the literature. We recall
for convenience the structure of these. Consider the $N$ fermion
density operator $\rho$ that commutes with the number operator
$\h{N}=\sum_{i=1}^{N}\h{c_i}^{\dg}\h{c_i}$. The site occupation basis
is
\beq |n_1,n_2,\ldots,n_N\kt=c_1^{\dg n_1}c_2^{\dg n_2}\dots c_N^{\dg n_N}
|0\kt
\eeq
where $n_i=0,1$ and $|0\kt$ is the vacuum. Note that there is an
isomorphism between these states and the states of $N$
qubits. Consider the reduced density matrix $\rho^R_{ij}$ of two sites
$i$ and $j$, where without loss of generality we can assume $i<j$. Due
to the restriction that $\sum_{i=1}^{N}n_i=m$, this operator has the
form:
\beq
\rho^R_{ij}=\left(\begin{array}{cccc}
        v_{ij}&0&0&0\\
        0& w_{1ij}&z_{ij}^{*}&0\\
        0&z_{ij}&w_{2ij}&0\\
        0&0&0&u_{ij} \end{array} \right).
\eeq

Here
\beqa
v_{ij}&=&<(1-\h{n}_i)(1-\h{n}_j)>\\
u_{ij}&=&<\h{n}_i\h{n}_j>\\
w_{1ij}&=&<(1-\h{n}_i)\h{n}_j>\\
w_{2ij}&=&<\h{n}_i(1-\h{n}_j)>\\
z_{ij}&=&<\h{c}_j^{\dg}\h{c}_i\prod_{l=i+1}^{j-1}(1-2\h{n}_l)>
\eeqa
and $<\h{A}>=\mbox{tr}(\h{A}\rho)$. The entanglement between the sites
(or qubits) $i$ and $j$ is measured here by the concurrence between them
that is given by
\beq
C_{ij}=C(\rho^{R}_{ij})=2 \mbox{max}(|z_{ij}|-\sqrt{u_{ij}v_{ij}},0).
\eeq

For the case $m=1$, the single-particle subspace, $u_{ij}=0$ and the
string of operators in the definition of $z_{ij}$ is not there.
If we write
$|l\kt=|0,\ldots,1_l,\ldots,0\kt$, a general one-particle
state is the superposition
\beq
|\alpha\kt=\sum_{l=1}^{N} \phi_l^{(\alpha)} |l\kt,
\eeq
where $\phi_l^{(\alpha)}=\br l|\alpha \kt$.
This then implies that the pairwise concurrence in this state are
\beq
C^{\alpha}_{ij}=2|\phi_i^{(\alpha)} \phi_j^{(\alpha)}|
\eeq

States that have large minimum pairwise concurrence can be said
to share entanglement better. As a gross but useful measure of
entanglement sharing we propose and study the average pairwise
concurrence in a given state. For single-particle states then:
\beq <C^{\alpha}>=\f{1}{d}\sum_{i<j}C^{\alpha}_{ij}=
\f{1}{d}\left( (\sum_{i=1}^{N}|\phi_i^{(\alpha)}|)^2-1\right),
\eeq where $d=N(N-1)/2$. From the structure of the average we see
that it has connections to measures of localization. In particular the
generalized entropies such as the Renyi entropy are related to the
averaged concurrence. For a given discrete probability
distribution$\{p_i,\,i=1,\ldots,N\}$, the Renyi entropy of order $q$
is defined as:
\beq
S^R_q[p]=\f{1}{1-q}\ln \sum_{i=1}^{N} p_i^q.
\eeq
This reduces to the usual information entropy as $q \rightarrow 1$.

Thus \beq <C^{\al}>=
\f{1}{d}\left(\exp(S^R_{1/2})-1\right) \eeq where $S^R_{1/2}$ is
the Renyi entropy of order one-half. Therefore we expect that
delocalized states share entanglement better, as an extreme case the
site localized state $|l\kt$ has zero average concurrence, as indeed
it is a completely separable state. For a study connecting the Renyi
entropy to localization we refer to \cite{Casati}. It must be noted
that we make this connection between localization and entanglement in
the case of one-particle states; it remains to be seen if there is
such a correspondence in the case of many-particle states.

We also note that $<C^{\al}>
\le 2/N$. This implies that for one particle states of qubits
there cannot be states whose minimum pairwise concurrence exceeds
$2/N$. This is the concurrence of isotropic states, which are defined
by identical pairwise density matrices. It is not yet known if the
above is true for states with larger number of particles
\cite{Woot02}.We go beyond the average and also study the 
{\it distribution} of concurrence, $p(C)$, in a given ensemble of
states, which will be representative of single states. In principle
then we can study various other averages of concurrence such as its
square etc., although we do not pursue this here.  

We show that for eigenstates of quantized classically chaotic systems,
the presence or absence of time-reversal symmetry, possibly a
generalized time-reversal, lead to very different
distributions. Near-zero concurrence are improbable for eigenstates of
time-reversal violating Hamiltonians, while they are most probable
otherwise. Time evolving states on the other hand, in either case,
behave as the eigenstates of time-reversal violating Hamiltonians. We
use, as a test model, the Harper Hamiltonian
\cite{Harper} (for a recent review and references, we point
to \cite{Artuso94}) which is an approximate model for
electrons in a two-dimensional crystal subjected to a perpendicular
magnetic field. This is a model with a rich spectral structure and a
metal-insulator transition that continues to be studied from various
viewpoints.

\section{Effect of onsite potentials}

\subsection{Integrable case}
In this subsection we study the effect of onsite potentials with a
view of also comparing an integrable situation to a nonintegrable one, a
more complex one to follow in the next section. We consider the Hamiltonian:
\beqa
H&=&\f{1}{2}\sum_{j=1}^N \h{c}_{j}^{\dg} \h{c}_{j+1}+ \f{g}{2}\sum_{k=1}^N 
\h{d}_{k}^{\dg} \h{d}_{k+1}+h.c.\nonumber \\&=& \sum_{j=1}^N [\f{1}{2}(\h{c}_{j}^{\dg} 
\h{c}_{j+1}+h.c.)+ g\cos(2\pi j/N) \h{c}_j^{\dg}\h{c}_j].
\eeqa
Here
\beq
\h{d}_k=\f{1}{\sqrt{N}}\sum_{j=1}^{N}\exp(2\pi i kj/N) \h{c}_j
\eeq
is the Fourier transform of the site annihilation operator
and $k$ is a momentum index. We will assume periodic boundary
conditions first: $\h{c}_{N+1}=\h{c}_1$, $\h{d}_{N+1}=\h{d}_1$.
$H$ is a one-dimensional Harper Hamiltonian with the onsite potential
being $\cos(2\pi q)$. We can think of the large-$N$ limit as approaching
a flow on the unit torus, with the classical Hamiltonian
\beq
\cos(2\pi p)+g \cos(2\pi q),
\eeq
and that we are considering its finite quantum mechanics with $N$
states.

We briefly indicate the reasoning involved. Note that the operators
\beq
\h{V}=\sum_{j=1}^{N} \h{c}_{j+1}^{\dg} \h{c}_{j},\;\;
\h{U}=\sum_{k=1}^{N} \h{d}_{k}^{\dg} \h{d}_{k+1}
\eeq
are unitary translation operators on the states $|l\kt$ and
$|k\kt\equiv\h{d}_k^{\dg}|0\kt$: $\h{V}|l\kt=|l+1\kt$ and $\br
k+1|=\br k|U$.  Thus the site and momentum states span a lattice on
the conventional unit torus phase space with the translation operators
$V$ and $U$ obeying a finite Weyl commutation relation; they are
discrete versions of $\exp(-i \h{p}a/\hbar)$ and $\exp(-i
\h{x}b/\hbar)$ (where $a,b$ are phase space shifts) respectively
\cite{ScBeSa}.  The torus-quantization implies the condition $\hbar=h/(2\pi)=
A/(2\pi N)$, where $A=1$ is the area of the unit torus phase space.
Also with $a=b=1/N$, as a lattice translation unit in phase space, and
with the eigenvalues of position and momentum being $l/N$ and $k/N$,
leads to the large-$N$ or classical Hamiltonian as specified above.

Thus we see that the original Hamiltonian is an integrable one in the
classical limit, as it has only a single degree of freedom.  From a
Bethe-Ansatz perspective the integrability of this Hamiltonian is
discussed in \cite{Wiegmann}.  We can also now easily visualize the
eigenstates of the Hamiltonian as being localized on the constant
energy curves of the classical Hamiltonian.  Thus although we cannot
solve the eigenvalue problem analytically we can understand the
features of all the states involved.

Another modification of the Hamiltonian is the class where the
the onsite potential is incommensurate with the lattice,  and
herein the Harper Hamiltonian shows a rich structure which has been studied
extensively. In particular we will modify the Hamiltonian to read
\beq
H=\sum_{j=1}^N [\f{1}{2}(\h{c}_{j}^{\dg} \h{c}_{j+1}+h.c.)+
g\cos(2\pi\sigma j/N) \h{c}_j^{\dg}\h{c}_j]
\eeq
where $\sigma$ is a real incommensurability parameter. For $\sigma/N$
a fixed irrational number (in the original Harper model, this is the
ratio of the flux through a lattice cell to one flux quantum) as $N$
tends to infinity a metal-insulator transition occurs at $g=1$ where
the spectrum is a Cantor set.

Firstly, the case $g=0$, $\sigma=1$ corresponds to an itinerant
particle on the lattice and the eigenfunctions are simply the momentum
states $|k\kt$. These clearly have pairwise concurrence $2/N$ for all
pairs and represent optimally-delocalized states in the site basis as
far as concurrence go. Due to double degeneracy however there exist
also eigenstates that have smaller entanglement.  For $g>0$
$\sigma=1$, the classical Hamiltonian above provides us the well-known
phase space of the Harper flow with two elliptic fixed points and two
hyperbolic fixed points per cell. From the Hamiltonian we know that
the energy eigenvalues are bounded by $-1-g \le E \le 1+g$.  The
classical phase space will consist of two separatrices, corresponding
to energies of the equilibrium points $(0,1/2)$ and $(1/2,0)$.  Thus
we know that for $0<g<1$ the quantum states with energies in the range
$-1+g<E<1-g$ are dominantly KAM rotational states spread along the
sites with the edge states being separatrix states.

There are states that will be localized in the site basis
corresponding to torus-quantized states around the elliptic fixed
points, while the hyperbolic orbits will provide the separatrix
states. When $g<1$ there are smooth phase-space curves along the
momentum direction and the separatrices localize states in momentum,
while at $g=1$ the two separatrices form a single diamond square and
for $g>1$ the separatrices tend to localize states along the position.
It is evident that as $g \rarrow
\infty$ there are states that are completely site localized. Thus the
classical picture also singles out $g=1$ as a special point.

\begin{figure}
\includegraphics[height=3in]{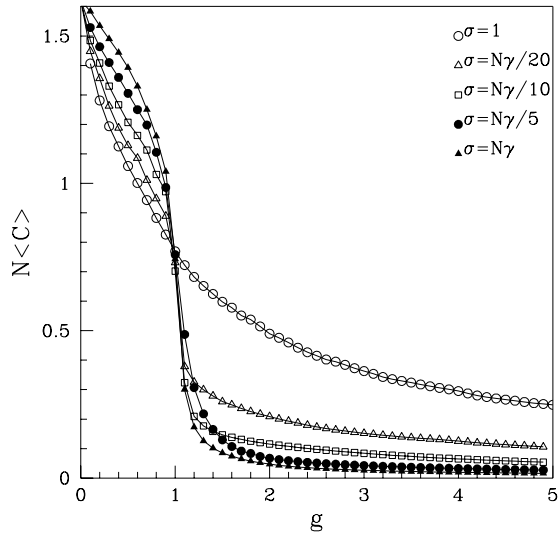}
\caption{Spectral averaged concurrence as a function of onsite potential
strength for the Harper Hamiltonian. $N=101$ and $\gamma=(\sqrt{5}-1)/2.$
\label{avc_vs_g_int}}
\end{figure}

Thus this elementary picture then indicates that as $g$
increases the average concurrence will tend to decrease. As a
further gross measure we average also over all the states, $\al$,
in the spectrum and show in Fig.~\ref{avc_vs_g_int} the decrease 
in the average
concurrence ($<C>$) as a function of $g$.
 Thus onsite potentials decrease concurrence as they tend to localize
 states. The point $g=0$ has exact double degeneracy and while the
 momentum states are eigenstates with the maximum concurrence of
 $2/N$, as soon as $g>0$ this degeneracy is broken and the states are
 continuation of combinations of the degenerate states with smaller
 entanglement. As an effect the apparent $g \rarrow 0$ limit is smaller
than $2$ in Fig.~\ref{avc_vs_g_int}.

In the same figure we also show the effect of $\sigma$. When $\sigma$
is an irrational number larger than unity the transition at $g=1$
becomes sharply visible. We note that the average concurrence
decreases dramatically as $g$ crosses unity, corresponding to a
metal-insulator transition in the infinite incommensurate chain.  This
is again a reflection of the fact that wavefunctions change from a
ballistic regime to an exponentially localized one. For rational
values of $\sigma$ close to these irrational values a transition is
still seen, due to finite $N$ effects. Thus apart from quantum phase
transitions, it is possible that the signature of entanglement will also be
present in metal-insulator transitions. 

\begin{figure}
\includegraphics[height=3in]{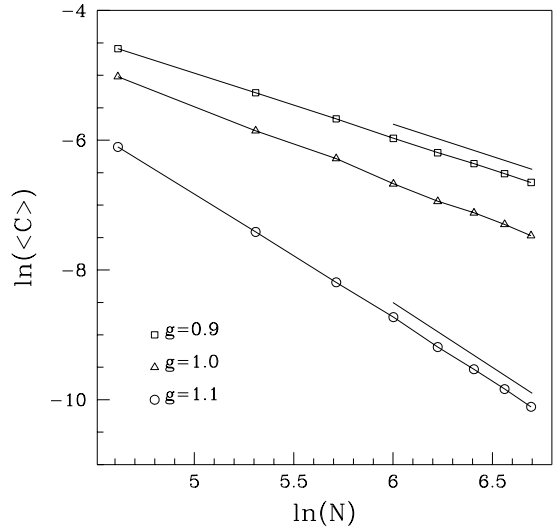}
\caption{Scaling of the spectral averaged concurrence with $N$ 
for $\sigma=N\gamma$, $\gamma=(\sqrt{5}-1)/2.$ Shown are three cases
corresponding to a metallic, critical and insulating regimes, the short
lines correspond to lines with slope $1$ $(g=0.9)$ and slope $2$ $(g=1.1)$.  
\label{lnc_vs_lnN_int}}
\end{figure}

Scaling behaviour with the size of the lattice $N$ is illustrated in 
Fig.~\ref{lnc_vs_lnN_int}, where a transition is seen from the scaling law
$<C> \sim N^{-1}$ for the metallic regime $g<1$ to $<C>\sim N^{-2}$ for 
the insulating regime $g>1$. 
The scaling in the localized or insulating regime is intuitively
reasonable, as there will be only a small number of significant
components, and hence only an order unity number of pairs can be
expected to have significant concurrence, hence the average will go
approximately as the inverse of the number of pairs. In the metallic
regime there is a more democratic spread of concurrence and results
in the $1/N$ scaling which we will later see to be the rule for
chaotic and random states. At the critical point, $g=1$, there does
not seem to be a simple good power law fit for the range of $N$
$(101-808)$ values used here, and while this warrants further study, we do 
not pursue this in this paper.

\begin{figure}
\includegraphics[height=3in]{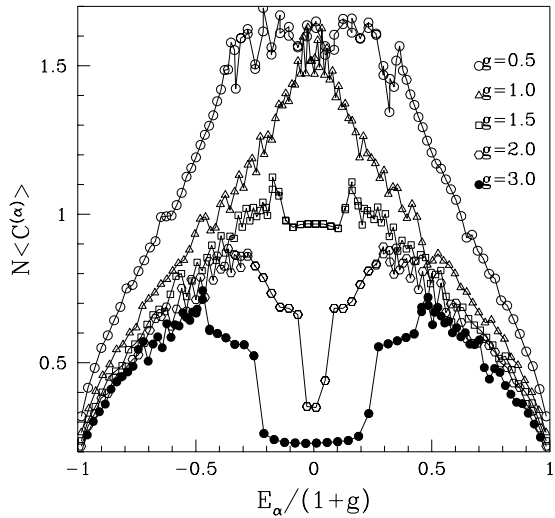}
\caption{Average concurrence in the individual states of the 
Harper Hamiltonian as a function of the energy, with $N=101$, $\sigma=1$ at various values of the onsite potential.} 
\label{avc_vs_enrg}
\end{figure}

In Fig.~\ref{avc_vs_enrg} is shown the average concurrence in
individual states of a given spectrum as a function of the 
energy of the state, appropriately scaled. For $g<1$ we see that there is
a plateau of large concurrence corresponding to states on the
rotational KAM invariant curves extending over all of the $q$ space,
while the edges are the separatrix states.  The tails on either side
correspond to states that are localized around the elliptic fixed
points and represent low-entanglement states on the average, having a
tendency to form coteries. As $g$ increases the plateau gets squeezed
out of existence and only the separatrix states remain at $g=1$. For
larger $g$ the invariant curves between the separatrices extend over
the momentum space rather the position and tend to start localizing in
the site basis. The very low concurrence states correspond to those
that are spread in momentum maximally and therefore highly site
localized.

\subsection{Nonintegrable Hamiltonians}

Nonintegrable Hamiltonians are the rule for systems with more than one
degree of freedom, or for many particle systems. While there are many
important interacting models in condensed matter physics such as the
Heisenberg model for which entanglement sharing has been studied, the
case of nonintegrability with the possibility of chaos has yet to be
explored.  We begin again with the simplest case of a single-particle
spectrum.  Building upon the Harper Hamiltonian which we have just
discussed, the kicked Harper Hamiltonian then provides us with a
suitable model.  The fact that we wish to remain on a one-dimensional
lattice means that we have to introduce a time-dependent onsite
potential to introduce nonintegrability. The kick-type of time
dependence leads to simple models that have been extensively studied
in the context of quantum chaos.  It has been pointed out that similar
models are of relevance in cyclotron resonance experiments in antidot
arrays \cite{IominFishman}.

\begin{figure*}
\label{phasespace}
\includegraphics[height=4in,width=4in ]{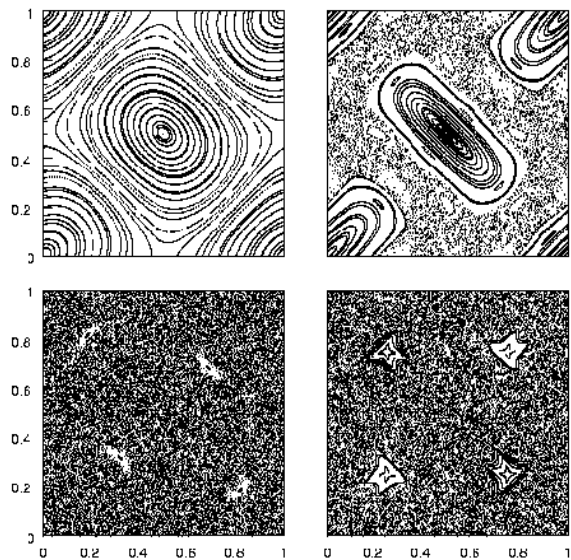}
\caption{The phase space $(q,p)$ of the classical map for $g=1$ and
$\tau=0.1$, $0.3$, $0.5$, and $0.7$ clockwise from top left.}
\end{figure*}

Thus the Hamiltonian we will consider is:
\beqa
H&=&\sum_{j=1}^N [\f{1}{2}(\h{c}_{j}^{\dg} \h{c}_{j+1}+h.c.)+
g\cos(2\pi j/N) \h{c}_j^{\dg}\h{c}_j \nonumber \\
&\times&\sum_{n=-\infty}^{\infty} \delta(2 \pi t/\tau-n)].
\eeqa
A train of impulses is provided at intervals of time $\tau/(2\pi)$. As
$\tau \rarrow 0$ we recover the integrable Harper equations. Note that
we have set for the nonintegrable case $\sigma=1$. The corresponding
large-$N$, classical Hamiltonian, is
\beq
H=\cos(2\pi p)+g \cos(2\pi q) \sum_{n=-\infty}^{\infty} \delta(2 \pi t/\tau-n)
\eeq
from which we get the canonical (area-preserving) map of the unit torus to
itself connecting phase-space variables immediately after two consecutive
impulses:
\beqa
q_{n+1}&=&q_n- \tau \sin(2\pi p_n)\nonumber\\
p_{n+1}&=&p_n + \tau g \sin(2 \pi q_{n+1}).
\eeqa
This map has been studied extensively and develops full fledged chaos
for large $\tau$ \cite{Leb90Lima91}. For completeness we illustrate
this transition to classical chaos in Fig.~\ref{phasespace}, fixing $g=1$ as in the
subsequent calculations too.

As is standard the Floquet operator (quantum map) connecting states
just after impulses is the quantum propagator:
\beq
\h{U}(\tau)= \exp(-i \tau g \cos(2 \pi \hat{q})/h)
\exp(-i \tau\cos(2 \pi \hat{p})/h).
\eeq
With $h=1/N$, we get the quantum version in the basis spanned by the
site-localized states $|l\kt$. The spectrum of the Floquet operator is
then of interest. We study the average concurrence and the
distribution of concurrence, in the eigenstates of the above quantum
map. The spectra of quantized chaotic systems are very sensitive to
whether time-reversal (TR) symmetry is present or not. However it was
seen in \cite{JayArul02} that the entanglement in pure states of
bipartite chaotic systems, consisting of Hilbert spaces of large
dimensions, is not sensitive to TR symmetry. This is because only the
density of eigenvalues of the reduced density matrix determines the
entanglement in the case studied there. We will see below that
concurrence sharing among many qubits is affected crucially by this
symmetry. To show this we change the boundary condition on the states
$|l\kt$ and introduce a phase, or equivalently change the boundary
conditions on the site creation operators:
\beq
|l+N\kt =\exp(-2 \pi i \beta)|l\kt; \;\; c_{l+N}^{\dg}=\exp(-2 \pi i \beta)
c_{l}^{\dg},
\eeq
where $ 0 <\beta<1/2$. This shifts the momentum eigenvalues to
$(k+\beta)/N$. We retain periodic boundary conditions on the momentum
states $|k\kt$ and note that in this kinematic framework the
momentum-site transformation is the Fourier transform
\beq
\h{d}_k=\f{1}{\sqrt{N}}\sum_{j=1}^{N}\exp(2\pi i (k+\beta)j/N) \h{c}_j.
\eeq
The phase $\beta$ is like a magnetic flux line threading the periodic
chain, which is a standard way to break the TR symmetry.

\begin{figure}
\includegraphics[height=3in]{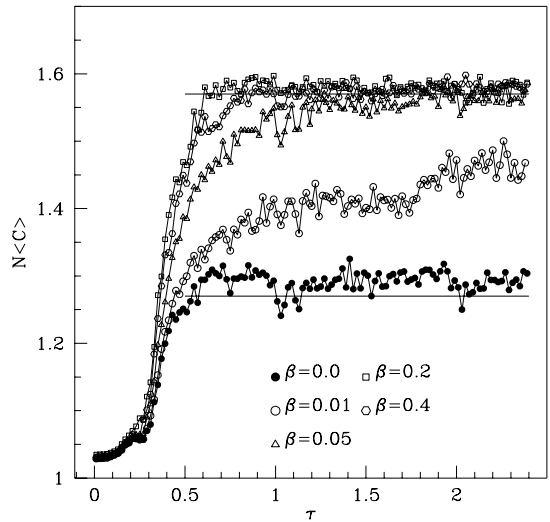}
\caption{Spectral averaged concurrence as a function of $\tau$ for the kicked  Harper Hamiltonian. Shown are both the TR symmetric ($\beta=0$) and several 
non-TR symmetric cases ($\beta \ne 0$), and $N=101$. The horizontal
lines correspond to the RMT averages $4/\pi$ and $\pi/2$
respectively.}
\label{avc_vs_tau_nint}
\end{figure}

In Fig.~\ref{avc_vs_tau_nint} the average concurrence is shown as a function of $\tau$ for
various values of the TR breaking phase $\beta$.  For any value of
this phase it is clear that along with a transition to classical chaos
there is an increase in the average pairwise concurrence and
corresponds to increasing delocalization of the states. The
concurrence sharing saturates after a transition to classical chaos,
at around $\tau=0.6$, and we expect that in this regime, RMT will be
able to model the concurrence, we show below that this expectation is
borne out.

It is also clear from this figure that breaking TR symmetry leads to
significantly larger entanglement sharing. The effect of time-reversal
breaking is pronounced in the chaotic regime as the delocalized states
experience the changed boundary conditions. In this figure we also
show how sensitive the TR breaking is by changing the phase only
slightly. For $\beta$ greater than the dimensionless Planck constant
$1/N$ we see a universal saturation effect. It is interesting that as
$\beta$ decreases the saturation seems to occur for larger $\tau$ which
is also a region of larger classical chaos.  We may conclude as a
general principle that entanglement sharing is more effective in
eigenstates of TR breaking Hamiltonians, and that this is a general
principle in the context of one-particle states is supported by an
analysis using RMT below.

We state results for two universality classes of RMT relevant here,
namely from the Gaussian unitary ensemble (GUE) for time-reversal 
breaking Hamiltonians, and the Gaussian orthogonal ensemble (GOE) for
TR preserving, spinless systems \cite{Haake,Mehta}.  We quantify the
above observations and note in advance, what we prove further on, that
the average concurrence calculated from RMT in the two cases are:
\beq <C^{\al}>= \left\{\begin{array}{ll}
     4/\pi N& \mbox{(GOE)}  \\
    \pi/ 2N& \mbox{(GUE)} \end{array} \right.
\label{avconc}
\eeq
The Fig.~\ref{avc_vs_tau_nint} saturation values agree well with these
estimates from RMT, which are shown as horizontal lines. However RMT
seems to predict a somewhat smaller value than the observed average in
the case when there is TR symmetry. Note that the RMT result used
above is a finite-$N$ exact value. Phase-space localization effects
on chaotic eigenfunctions due to classical periodic and homoclinic
orbits do lead to significant deviations from RMT
\cite{ArulNickSteve}, however here it is interesting that the
deviations seem to be more when there is TR symmetry. Concurrence
could be a sensitive measure to study deviations of real
eigenfunctions of quantized chaotic systems from RMT predictions;
further study on this aspect is required.

The average concurrence promises in the case of the one-particle
spectrum to be an interesting measure of localization.  We also
emphasize that these results are only dependent on the single-particle
nature of the states and are {\it independent} of the dimensionality,
although our models are one-dimensional. Calculations with different
$N$ not presented here also confirm further the RMT scaling $<C>\sim
N^{-1}$, which we also observed in the metallic regime of the
integrable Harper Hamiltonian, and as in that case we postpone a more
extensive scaling analysis.

\begin{figure}
\includegraphics[height=3in]{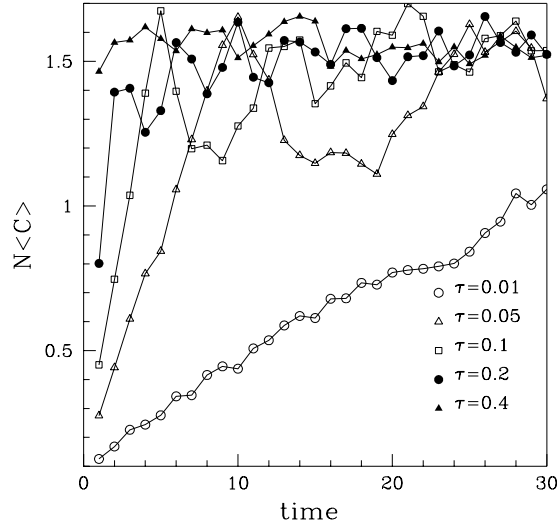}
\caption{Average concurrence for a nonstationary state (initially $|l=21\kt$),
as a function of time. $N=101$, and  near integrable to chaotic cases
are shown.}
\label{avc_vs_time_nint}
\end{figure}

Time evolution intrinsically involves complex vectors and
therefore we expect that time-evolving states will share,
under a quantum chaotic evolution, entanglement that is identical
with that of TR breaking Hamiltonian eigenstates. This is borne
out in Fig.~\ref{avc_vs_time_nint} where several cases ranging from near integrable to
chaotic are shown. The near-linear increase of the average
concurrence in time for near-integrable systems is replaced by a
rapid increase to the TR breaking average of $\pi/2N$ around which
there are small fluctuations. The initial state in all these cases
is a site-localized one with null entanglement.
While the average pairwise concurrence of chaotic eigenstates is
larger than that of regular states, it is reasonable to expect the
opposite if one were to only include near neighbor pairs of sites. We
expect that the nearest neighbors are treated preferentially in
regular states while for random or chaotic states the connections from
one site to another is also random. This expectation comes from the
fact that regular states are smoother than the more fragmented
structures one finds on a coarse scale for chaotic states. This is in turn
related to tori quantization of regular states, as opposed to a situation 
more akin to superposition of random waves for chaotic states \cite{Gutz}.

\begin{figure}
\includegraphics[height=3in]{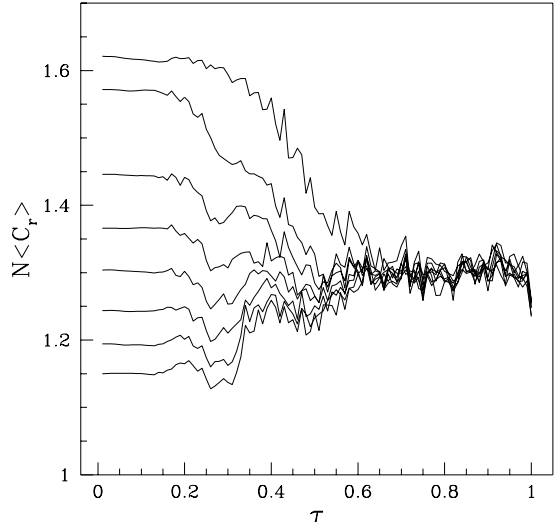}
\caption{Average $r$-th neighbor concurrence as function of $\tau$
for the kicked Harper Hamiltonian.  From top to bottom $r=1$ to $15$
in steps of $2$ and $N=101$ in all cases.}
\label{avcr_vs_tau_nint}
\end{figure}

\begin{figure}
\includegraphics[height=3in]{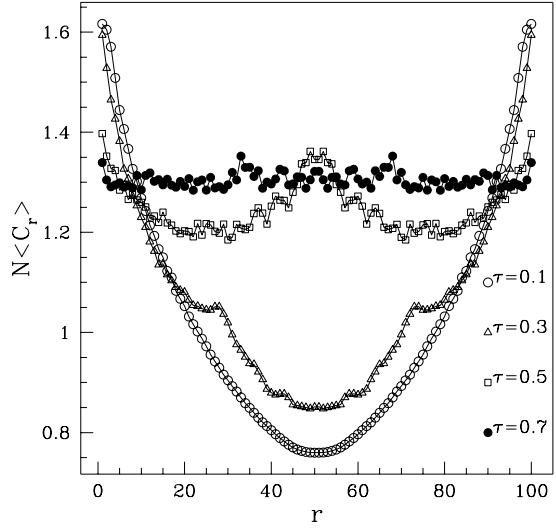}
\caption{Average $r$-th neighbor concurrence as function of $r$
for the kicked Harper Hamiltonian. Shown are cases ranging from the 
near integrable to the chaotic. $N=101$, and $\beta=0$.}
\label{avcr_vs_r_nint}
\end{figure}

Thus we define the $r$-th neighbor
average concurrence:
\beq C^{(\al)}_r=\f{1}{N}\sum_{i=1}^{N}C^{(\al)}_{ii+r} \eeq
In Fig.~\ref{avcr_vs_tau_nint} this is shown, after averaging over the spectrum $\al$,
for various $r$ as a function of $\tau$
for TR symmetric eigenstates. It is clear that the correlation between
near-neighbor pairs is indeed much stronger for regular states.  There
is a correlation length beyond which the entanglement falls below that
of the random or chaotic states average of $4/\pi N$. This correlation
length is then an interesting quantum length scale of the problem; we
call this a quantum scale as it refers to the intrinsically quantum
property of entanglement. In Fig.~\ref{avcr_vs_r_nint} we show how $C_r^{(\al)}$ falls as
a function of $r$ for various $\tau$, after averaging over the
complete spectrum $\{\al\}$.

\subsection{RMT and concurrence}

In this section we derive the averages stated and demonstrated above,
as well as the {\it distributions} of the concurrence between sites of
one-particle states using random matrices as models. The
eigenfunction component distributions are derived within RMT by
invoking a microcanonical distribution with the constraint being
normalization. If ${x_1,x_2,...x_d}$ are real numbers distributed
uniformly over the $d$ dimensional spherical surface of unit radius
(normalization), the reduced density of $l$ variables is given by
\cite{Haake}:
\beqa
P^{(d,l)}(x_1,x_2,\ldots,x_l)&=&\pi^{-l/2} \f{\Gamma(d/2)}{\Gamma((d-l)/2)}\nonumber \\
&\times&\left(1-\sum_{n=1}^{l} x_n^2 \right)^{(d-l-2)/2}.
\eeqa

In the case of TR symmetric systems, there exist time-reversal adapted
bases wherein the components of the eigenfunctions are real and these
may then be taken to be the $x_i$ above with $d=N$. In the case of TR
breaking Hamiltonians there are no such bases and the eigenfunctions
are generically complex, in which case we identify the real and
imaginary parts of the state components with the $x_i$ and $d=2N$
\cite{Haake}. Thus the average concurrence for the GOE case, relevant for TR symmetric Hamiltonians, may be calculated as the integral:
\beq
<C>=\int_{R_2} dx_1 dx_2 \,P^{(N,2)}(x_1,x_2)\, 2\sqrt{x_1^2
x_2^2}=4/\pi N.
\eeq
The region $R_2$ is the interior of the circle $x_1^2+x_2^2 \leq 1$.
For the GUE case, where the wavefunctions have complex components and
$d=2N$, the average concurrence is:
\beqa
<C>&=&\int_{ R_4}
dx_1 dx_2 dx_3 dx_4\, P^{(2N,4)}(x_1,x_2,x_3,x_4)\nonumber \\
&\times& 2\sqrt{x_1^2 +x_2^2}
\sqrt{x_3^2 +x_4^2} =\pi/2 N.
\eeqa
The region $R_4$ is now the 4-sphere volume:$x_1^2+x_2^2+x_3^2+x_4^2\leq1$.
These are the formulae stated in Eq. (\ref{avconc}).

In order to calculate the distributions themselves we choose to use
the large $N$ forms of the distributions, when the components tend to
become independent. Let $\rho(x)$ be a single-component distribution
of $x=|\phi^{(\alpha)}_j|^2$.  The distribution function, $\rho(x)$,
is known to be different for the two universality classes used here.
The GOE distribution, the Porter-Thomas distribution, was first used
in the study of nuclear resonance widths \cite{RMTrev}.

\beq \rho(x)= \left\{\begin{array}{ll}
     \sqrt{N/2\pi x}\;\exp(-Nx/2)& \mbox{(GOE)}  \\
    N \;\exp(-Nx)& \mbox{(GUE).} \end{array} \right.
\eeq

Thus the concurrence distribution, $p(C)$, is then straightforward
to calculate for one particle states. We state the distributions
for the scaled concurrence $c=N\, C$:

\beq p(c)=\int_{0}^{\infty} \int_{0}^{\infty}\,
\delta(c-2\,N\sqrt{xy}) \,\rho(x)\rho(y)\, dx dy. \eeq The result
is: \beq p(c)= \left\{\begin{array}{ll}
     (1/\pi)K_0(c/2)& \mbox{(GOE)}  \\
    c K_0(c)& \mbox{(GUE),} \end{array} \right.
\eeq where $K_0$ is the modified Bessel function which has a
logarithmic divergence at the origin. The average concurrence, stated
in Eq.~(\ref{avconc}) and derived above, also follow from the single
component distributions, {\it i.e.},  at least in the
averages there are no corrections coming from correlations between the
components.
\begin{figure}
\includegraphics[height=3in]{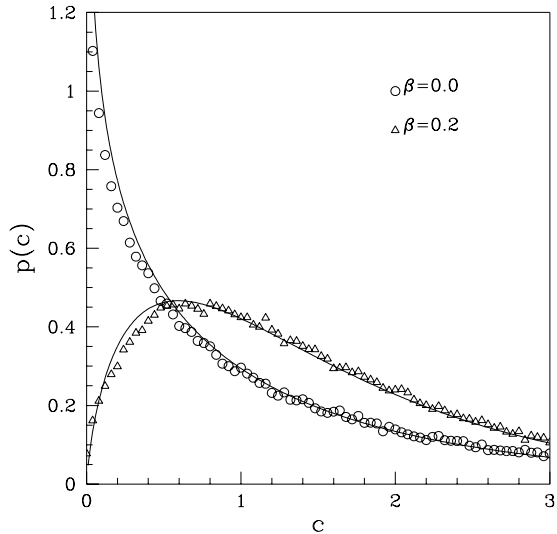}
\caption{The concurrence distributions for the kicked Harper
Hamiltonian with $\tau=0.8$ and $N=101$. Shown using points 
are the cases of time reversal preserving ($\beta=0$) and
time reversal breaking ($\beta=0.2$) Hamiltonians. The smooth 
curves are the RMT predicted distributions.}
\label{distrib}
\end{figure}

We recall that one-particle states that maximally share entanglement
are those whose reduced density matrices for all the pairs are
identical, such as the non-interacting case eigenstates states
$\phi_j^{(k)}=\exp(2 \pi ij k/N)/\sqrt{N}$. If we take the pairwise
concurrence, $2/N$ in this case to be a marker, the fraction of pairs
with concurrence larger than this is 
\beq \int_{2}^{\infty}p(c)\,
dc\,=\,.21,\, .28\,
\;\;\mbox{for GOE, GUE, resp.}\eeq Thus a significant proportion of
the pairwise concurrence in a one-particle random state is higher than
$2/N$. For large $c$ (practically greater than $2$) the asymptotic
distributions are:
\beq p(c) \sim e^{-c/2}/\sqrt{\pi c},\;\; \sqrt{\pi c/2}
\,e^{-c}\eeq for the two cases of GOE and GUE respectively.

In Fig.~\ref{distrib} we compare the distributions from RMT with numerical
calculations. To do this we combine the pairwise concurrence of all
the eigenstates into a concurrence ensemble. We see that there is
excellent agreement between the theory and numerical calculations,
although there are discernible deviations for small concurrence. It
must be noted that these distributions are {\it universal}, they are
independent of system except for the requirement of a classically chaotic
limit. Thus it is clear that TR symmetry could play a crucial role in
the way entanglement is shared in a quantum state.  Also the case of
the Gaussian symplectic ensemble has not been considered here due to
the additional complexity of a Kramer's degeneracy.

\section{Conclusions}

We have studied entanglement sharing in one-particle states using the
measure of concurrence. We have studied the average of the concurrence,
spectral averaged concurrence, as well as the distribution of
concurrence in a given state. Our attempt has been to begin studying
the effect of non-integrability and chaos in this interesting quantum
measure that has possible applications in quantum information theory.
We have found that chaos in the corresponding quantum system implies
that concurrence in individual one-particle states are distributed in 
an universal manner that depends only on the presence or absence of 
TR symmetry. The absence of TR symmetry has been shown to lead to 
more entanglement sharing, and we have quantified these statements
with the help of RMT. 

Also a transition to chaos has been shown to accompany an increase in
the spectral averaged concurrence. We expect that spectral averaging
will have small effect in the case of chaotic systems, while there
will be much larger state to state fluctuations in the case of regular
states. We have also made connections between the averaged concurrence
and measures of state localization such as the Renyi entropy,
something that enables us to qualitatively understand the behaviour of
entanglement sharing in these states. For instance we have
demonstrated that for regular states, near-neighbor concurrence is
preferred, while for chaotic states there is no such metric
preference. Due to the essential simplicity of the concurrence in
one-particle states, we have been able to analyze details such as
concurrence distributions. It is of much interest to see how many of
the conclusions, for instance that concerning the role of TR symmetry,
carry over to general states.

The effect of onsite integrable potentials has also been studied and
it is noted that transitions such as metal-insulator transitions are
reflected in the way entanglement is shared. It is shown that in the
metallic regime entanglement is shared better, while in the insulating
regime, it is not. This is reasonable due to the connections between
wavefunction localization and concurrence noted above. Also the
finite-size scaling of the concurrence in the metallic and insulating
cases have been noted. A more detailed scaling analysis in all cases,
including the critical point is warranted. The hypothesis that quantum
information theory will be an approach to study ``complex'' quantum
systems \cite{Nielsencomplex} is borne out here in the context of
entanglement and single-particle chaos. The way in which
many-particle states differ are significant and ongoing work on this
will soon be reported.

\begin{acknowledgments}
AL thanks Prof. M. K. Verma and other members of the 
Dept. of Physics, IIT Kanpur, for their invitation and wonderful 
hospitality during
his sabbatical at which this work was done.
\end{acknowledgments}

\end{document}